\documentclass[10pt,letterpaper,twocolumn]{article} 

\usepackage{ol2}
\usepackage{graphicx}
\usepackage[draft]{hyperref}
\usepackage{amsmath}

\newcommand  {\dip}   {\mathrm{dip}}
\newcommand  {\pc}    {\mathrm{pc}}

\def\figurename{\textbf{Figure}}
\def\@caption@fignum@sep{$|$ }   

\makeatletter
\renewcommand{\fnum@figure}{{\small{\bf \figurename\ \thefigure}}}
\makeatother

\newcommand {\Tubingen} {Physikalisches Institut, Eberhard-Karls-Universit\"{a}t T\"{u}bingen, Auf der Morgenstelle 14, D-72076 T\"{u}bingen, Germany}
\newcommand {\Brazil} {Instituto de F\'isica de S\~ao Carlos, Universidade de S\~ao Paulo, 13560-970 S\~ao Carlos, SP, Brazil}

\begin{document}

\twocolumn[ 

\title{Optical parametric oscillation with distributed feedback in cold atoms}

\author{Alexander Schilke$^{1}$, Claus Zimmermann$^1$, Philippe W. Courteille$^2$ and William Guerin$^{1,*}$}

\address{
$^1$\Tubingen \\
$^2$\Brazil \\
$^*$e-mail: william.guerin@pit.uni-tuebingen.de
}

\ \\

]

\small

\noindent \textbf{There is currently a strong interest in mirrorless lasing systems \cite{JOpt2010_SpecialIssue}, in which the electromagnetic feedback is provided either by disorder (multiple scattering in the gain medium) or by order (multiple Bragg reflection). These mechanisms correspond, respectively, to random lasers \cite{Wiersma:2008} and photonic crystal lasers \cite{Noda:2010}. The crossover regime between order and disorder, or correlated disorder, has also been investigated with some success \cite{Conti:2008,Mahler:2010,Noh:2011}.
Here, we report one-dimensional photonic-crystal lasing (that is, distributed feedback lasing \cite{Kogelnik:1971,Yariv:book}) with a cold atom cloud that simultaneously provides both gain and feedback. The atoms are trapped in a one-dimensional lattice, producing a density modulation that creates a strong Bragg reflection with a small angle of incidence. Pumping the atoms with auxiliary beams induces four-wave mixing, which provides parametric gain. The combination of both ingredients generates a mirrorless parametric oscillation with a conical output emission, the apex angle of which is tunable with the lattice periodicity.}

Among the possible systems that can be used to produce and study mirrorless lasers, cold atoms are interesting because of their specific properties that differ from these of standard photonic materials. First, they are resonant point-like scatterers, producing extremely narrow spectral features (gain curves, scattering cross-section), which can provide flexibility \cite{Gottardo:2008} or give new effects. Second, their temperature is low enough to make Doppler broadening negligible in most situations, but large enough to make them move substantially on a millisecond timescale, which makes disorder-configuration averaging or dynamic evolution from order to disorder very easy. Third, cold atoms are well isolated from the environment, which makes them good candidates in the search for quantum effects.

Conventional lasing has already been demonstrated when cold atoms are used as the gain medium \cite{Guerin:2008,Vrijsen:2011}, as has radiation trapping due to multiple scattering \cite{Labeyrie:2003}, and efforts are under way to combine both factors to obtain random lasing \cite{Froufe:2009,Guerin:2010}. In the opposite regime, a one-dimensional photonic bandgap (PBG), yielding efficient Bragg reflection of light, has recently been demonstrated in a cold, ordered atomic vapour \cite{Schilke:2011}. In this Letter, we demonstrate optical parametric oscillation (OPO) with distributed feedback (DFB) in cold atoms trapped in a one-dimensional optical lattice, by combining the PBG with four-wave mixing (FWM), which provides the gain mechanism \cite{Guerin:2008,Abrams:1978,Leite:1986,Pinard:1986}.

\begin{figure}[b]
\centerline{\includegraphics{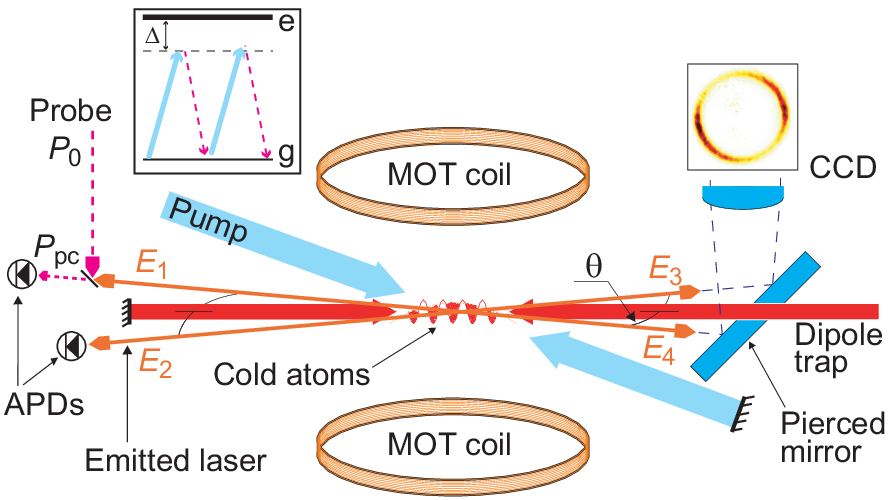}}
\caption{\footnotesize\textbf{Schematics of the set-up}. Cold atoms are trapped in the lattice formed by a retroreflected dipole trap. The pump beam is also retroreflected and has an incident angle of $\sim 8^\circ$. Above threshold, the system emits light with an angle $\theta$ around the lattice: in any given plane including the lattice, four waves are coupled. $E_1$ and $E_4$ as well as $E_2$ and $E_3$ are coupled by the phase-conjugation process; $E_1$ and $E_3$ as well as $E_2$ and $E_4$ are coupled by the Bragg reflection. The emitted light is detected by avalanche photodiodes (APDs) and the beam cross section is observed by a charge-coupled device (CCD) camera. Additionally, a probe beam (incident power $P_0$) can be used to measure the phase-conjugate reflectivity $R_\pc = P_\pc/P_0$, where $P_\pc$ is the reflected power. Inset: scheme of the four-photon transition corresponding to FWM. Upward (downward) arrows represent pump (probe and conjugate) photons. MOT, magneto-optical trap.}\label{fig.setup}
\end{figure}

\begin{figure*}[t]
\centerline{\includegraphics[width=14cm]{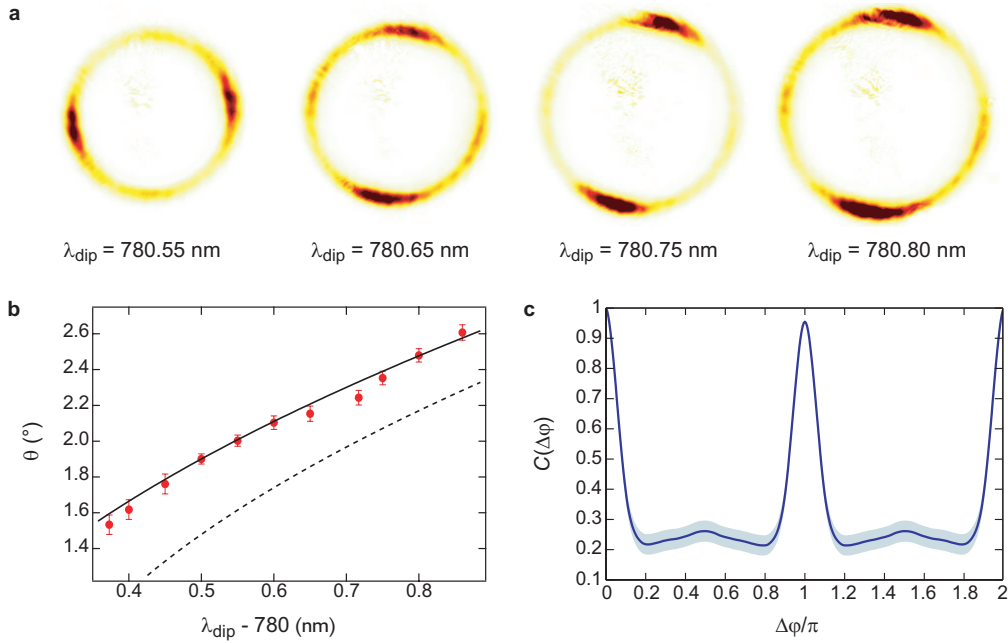}}
\caption{\footnotesize\textbf{Transverse mode of the emitted beam. a,} Cross-section of the beam for different $\lambda_\dip$. Each image is the result of an average of about 100 pictures of integration time $200~\mu$s each. \textbf{b,} Measured emission angle $\theta$ as a function of $\lambda_\dip$ (red symbols). Errors bars indicate the uncertainty in fitting the radius on images and the angle calibration (see Methods). The dashed line is the Bragg condition (equation \ref{eq.Bragg}) with $\bar{n} = 1$ and the solid line is a fit with $\bar{n}$ as a free parameter. \textbf{c,} Angular correlation function $\mathcal{C}(\Delta\varphi)$ computed from the $\lambda_\dip = 780.75$~nm data. The solid line is the average over all images and the blue area repressents the statistical uncertainty.}\label{fig.profile}
\end{figure*}

We trapped cold $^{87}$Rb atoms in a one-dimensional lattice of tunable wavelength $\lambda_\dip$. The trapping beam is retroreflected to generate a potential of periodicity $\lambda_\dip/2$ (Fig.\ \ref{fig.setup} and Methods). Typically, $N = 5\times 10^7$ trapped atoms were distributed over a length $L \approx 3$~mm ($\sim 7,700$ atomic layers) at a temperature $T \approx 100\, \mu$K, leading to a root-mean-square (r.m.s.) transverse radius of the cloud $\sigma_\perp \approx 60\, \mu$m.
Such an atomic pattern gives rise to a periodic modulation of the refractive index $n$ and we have shown recently \cite{Schilke:2011} that the very small modulation amplitude $\Delta n \approx 10^{-3}$ inherent to dilute vapours (density $\rho \approx 10^{12}$~cm$^{-3}$) could be balanced by the large number of layers, provided that the Bragg condition is fulfilled for a small angle of incidence and for a frequency slightly off the atomic resonance to avoid too much scattering losses. A Bragg reflection as efficient as 80\% can be obtained.

In the present experiment, we investigated the situation when \emph{gain} was added to the system. Cold atoms can amplify light when pumped by auxiliary near-resonant beams, and several gain mechanisms have already been demonstrated (see ref. \cite{Guerin:2008} and references therein). The combination of enough gain and multiple Bragg reflection should lead to DFB lasing.

However, in this system, the stability of the feedback mechanism is a critical issue. We trap the atoms using a lattice beam that is far red-detuned from the atomic transition (D2 line of $^{87}$Rb, wavelength $\lambda_0 = 780.24$~nm in vacuum, linewidth $\Gamma/2\pi=6.07$~MHz), so that the Bragg condition can only be fulfilled for a non-zero propagation angle $\theta$, relatively to the lattice axis, given by
\begin{equation}\label{eq.Bragg}
\lambda_\dip \cos \theta = \lambda_0/\bar{n}
\end{equation}
where $\bar{n}$ is the averaged refractive index of the atomic medium.
Because of this angle, typically $\theta \approx 2^\circ$ for $\lambda_\dip = 780.6$~nm, the light beam is not perfectly reflected onto itself such that it transversely walks off and leaves the interaction volume after a certain number of Bragg reflections.

Because this walk-off effect plays an important role, we must consider, for each plane containing the lattice, four different waves having a propagation angle $\theta$ relatively to the lattice axis (Fig. \ref{fig.setup}). In that case, a `standard' gain amplifies one wave (e.g. $E_4$ in Fig. \ref{fig.setup}) while the Bragg reflection couples it to another one ($E_2$) with a walk-off. This problem can be overcome by producing gain with a phase-conjugation mechanism such as degenerate or nearly degenerate FWM \cite{Abrams:1978,Leite:1986,Pinard:1986,Guerin:2008}. Then, each wave generates a backward phase-conjugated wave. Combined with the Bragg reflection, this leads to a global, walk-off free coupling between all four waves (Fig. \ref{fig.setup}), which favours an oscillatory behaviour.

Experimentally, inducing FWM in a phase-conjugation configuration is done by simply retroreflecting a near-resonant, linearly-polarized pump beam (Fig.\ \ref{fig.setup}). Its typical detuning from the $F\!=\!2\!\rightarrow\!F'\!=\!3$ closed transition is $\Delta\!=\!-5 \Gamma$, it makes an incident angle with the lattice of $\sim 8^\circ$ and it is collimated with a waist $w \simeq 2.4$~mm, thus ensuring a nearly-homogeneous pumping of the whole lattice. To avoid optical pumping into the dark hyperfine $F\!=\!1$ ground state, a repumping laser is kept on all the time. In addition, a weak probe beam, phase-locked with the pump, can be used for pump-probe experiments or as a local oscillator.

\begin{figure}[t]
\centerline{\includegraphics[width=8cm]{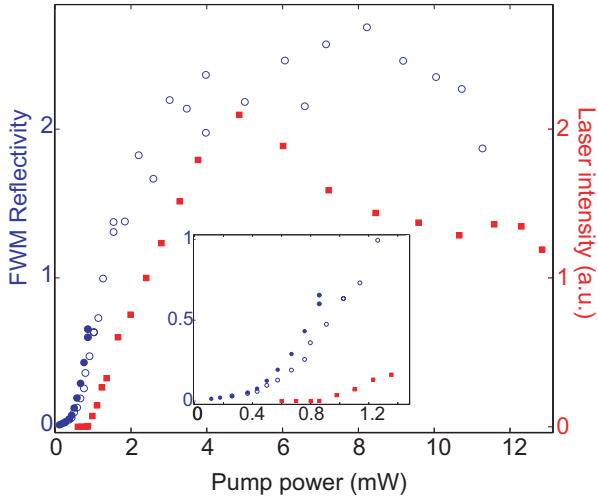}}
\caption{\footnotesize\textbf{Emitted laser power and maximum FWM reflectivity as a function of the pump power.} Red filled squares: laser intensity. Blue filled circles: FWM reflectivity below the threshold. The pump detuning is $\Delta = -5 \Gamma$. Blue open circles: FWM reflectivity measured without lattice (to compensate the lattice-induced light-shift, we increased the pump detuning to $\Delta = -7 \Gamma$). Inset: zoom-in for low pump power. The comparison between the reflectivity measured with and without the lattice shows that the lattice increases the reflectivity when approaching the threshold, which demonstrates a positive feedback effect. Above threshold, only the reflectivity without lattice can be measured.}\label{fig.FWM_Laser_vs_Ppump}
\end{figure}

When the pump power $P$ overcomes a certain threshold, namely $P_\mathrm{th} \approx 1$~mW for $\Delta=-5\Gamma$ (Fig. \ref{fig.FWM_Laser_vs_Ppump}), we observe a strong, directional light emission that can be recorded either with avalanche photodiodes or with a charge-coupled device (CCD) camera (Fig. \ref{fig.setup}). This radiation is due to OPO with DFB in the cold atom sample, with FWM as the gain mechanism. This interpretation is supported by many observations, as discussed in the following.

First, this radiation is obtained only with a retroreflected pump beam, whose alignment is critical, which is a strong indication that FWM is at work (with only one pump beam, we observe a strong Raman gain \cite{Guerin:2008} in pump-probe experiments but no laser). Second, the polarization of the emitted radiation is linear and orthogonal to the one of the pump beam. This is also consistent with the properties of FWM, which is much more efficient for orthogonal pump and probe polarizations \cite{Lezama:2000}. Third, we measured the frequency of the emitted light by a beat note experiment (see Supplementary Information). The emitted light is just a few kilohertz detuned from the pump frequency, which is consistent with nearly degenerate FWM \cite{Vallet:1990} and inconsistent with Raman gain, for which the amplification line was detuned by $\sim 200$~kHz with similar parameters.

The role of Bragg reflection as the feedback mechanism is demonstrated by the beam shape. Emission occurs with angle $\theta$ from the lattice axis and, because of axial symmetry, forms two cone-shaped beams on each side of the lattice.
From images of the beam cross-section (Fig.\ \ref{fig.profile}a), we extract the emission angle $\theta$ as a function of $\lambda_\dip$. A fit with $\bar{n}$ as a free parameter is fully consistent with the Bragg condition (equation \ref{eq.Bragg}) and gives $\bar{n}-1 = (2.2 \pm 0.5) \times 10^{-4}$ (Fig.\ \ref{fig.profile}b), which is the expected order of magnitude given the atomic density (the refractive index depends also on the pump power).

The intensity profile $I(\varphi)$, where $\varphi$ is the in-plane angle, exhibits strong shot-to-shot fluctuations. However, there is always a symmetry, that is $I(\varphi) \approx I(\varphi+\pi)$. This can be more precisely quantified by computing from the images the angular correlation function $\mathcal{C}(\Delta\varphi) = \langle I(\varphi) I(\varphi + \Delta\varphi) \rangle /\langle I(\varphi)^2 \rangle$, which shows indeed a very strong correlation $\mathcal{C}(\pi) \approx 0.96$ (Fig.\ \ref{fig.profile}c). This correlation comes from the couplings, namely the Bragg reflection and the phase conjugation, that exist in any given plane between the four directions of emission (Fig. \ref{fig.setup}). In contrast, the shot-to-shot fluctuations can be understood by the random direction of the initial spontaneous emission event that triggers the laser oscillation and by the absence of coupling between waves in different planes.
In addition, we note that even after averaging, the intensity is not uniformly distributed along the ring (Fig.\ \ref{fig.profile}a), indicating that some effects break the axial symmetry (the incident angle of the pump beam and possibly some residual astigmatism in the lattice beam).

Finally, we have measured the emitted intensity as a function of the pump parameters.
We observed that the OPO is not very sensitive to the precise value of the pump detuning $\Delta$, provided it is negative (red-detuned) and not too close to resonance: we did not observe any significant change between $\Delta = -4 \Gamma$ and $\Delta = -10 \Gamma$. Closer to resonance or with a positive detuning, the pump-induced heating destroys the lattice, and the laser emission is inhibited or much weaker.
The dependance on pump power is much more important. Above the threshold, the emitted power increases with pump power up to an optimum at $P \approx 5$~mW, above which a decrease is observed (Fig.\ \ref{fig.FWM_Laser_vs_Ppump}). We estimate the maximum emitted power to be $\sim 3~\mu$W on each side.
The decrease of the reflectivity past an optimum pump intensity is a known behaviour of FWM \cite{Abrams:1978}. In our case, we also suspect a detrimental mechanical effect (heating or residual radiation pressure) which destroys the lattice. We indeed observed, in the temporal behaviour of the emitted light, that radiation is sustained for a longer time for lower pump intensities, the longest duration being $\sim 0.5$~ms.

In summary, it is interesting to note that the threshold condition requires $\sim 50\%$ phase-conjugate reflectivity measured with a disordered sample (Fig. \ref{fig.FWM_Laser_vs_Ppump}) and $\sim 80\%$ Bragg reflectivity measured with a passive sample \cite{Schilke:2011}, but when both
ingredients are combined, these quantities are modified and we can no longer determine the amounts of feedback and gain, in contrast to standard lasers, in which the gain medium and cavity can be characterized independently of one another. This is also what makes our system original and interesting.

To conclude, let us discuss the possible applications of this DFB OPO.
First, we stress that the idea of combining FWM with a red-detuned DFB grating to generate conical beams (and possibly Bessel beams \cite{McGloin:2004}) might also be applied in other systems, such as semiconductors \cite{Inoue:1987,Mecozzi:1993}.
In contrast, on-axis feedback could be obtained by using higher-order Bragg reflection, for exemple, with $\lambda_\dip = 2\lambda_0/\bar{n}$. Combining this idea with frequency up conversion schemes \cite{Zibrov:2002,Schultz:2009,Akulshin:2009,Vernier:2010}, one might be able to produce mirrorless oscillation at $\lambda_0$ with pumps at longer wavelengths. In principle, $\lambda_0$ can be any atomic transition, possibly in the ultraviolet range.
Another possible application is the generation of `twin beams', that is, pairs of beams with a relative intensity noise below the standard quantum limit. It is indeed well known that FWM induces quantum correlations and it has been shown that FWM oscillators above threshold could create twin beams \cite{Vallet:1990}. We have already demonstrated strong angular classical correlations, and more work is needed to investigate, theoretically and experimentally, the question of possible quantum correlations in such systems. Here, the correlations should concern the four directions of emission in each plane.
Finally, extension to a three-dimensional geometry, using the bandgap predicted in ref. \cite{Antezza:2009}, is an exciting open question.
\\

\noindent\textbf{Methods}

\begin{scriptsize}
\noindent \textbf{Atomic sample preparation and experimental cycle.} The atoms were  trapped and cooled in a magneto-optical trap (MOT) loaded from a background vapour in 1~s. Stages of compression and molasses were used to increase the density and decrease the temperature, before loading the atoms in the dipole trap, which was generated by a home-made continuous-wave Ti:sapphire laser, following the design of ref. \cite{Zimmermann:1995}. The maximum available power is $1.3$~W and its wavelength $\lambda_\dip$ was tunable in the range 770--820~nm. The beam was focussed with a waist ($1/e^2$ radius) $w_\dip = 220\,\mu$m at the MOT position (Rayleigh length $z_\mathrm{R} \approx 20$~cm). The lattice was then formed by retroreflecting the beam. After loading the optical lattice, the molasses beams were switched off and a waiting time of 20~ms allowed the untrapped atoms to escape. Then, we could either characterize the trapped sample by absorption imaging, perform pump-probe spectroscopy to measure transmission, Bragg reflection and phase-conjugate reflection spectra, or shine only the pump beams to observe the emitted light. When the phase-conjugate reflectivity was measured without lattice (Fig. \ref{fig.FWM_Laser_vs_Ppump}), the lattice was switched off with a mechanical shutter in $130~\mu$s and the atoms freely expanded for 1 ms before measurements. This time of flight was large enough to completely smooth out the ordered structure and small enough to keep the optical thickness constant. All the stages following the initial MOT loading lasted only a few milliseconds, so that the total cycle duration was not much longer than 1~s. The repetition rate was thus $\sim 1$~Hz, which allowed quick averaging over many realizations.\\

\noindent \textbf{Detection tools.} To measure the phase-conjugate reflectivity, we used APDs and obtained spectra by sweeping the probe frequency with an acousto-optic modulator in double-pass configuration. Because the probe power fluctuated during the sweep, the recorded reflected intensity was divided by a reference intensity recorded simultaneously with another APD illuminated by a part of the probe beam. The relative sensitivity of both APDs was calibrated with a 10\% uncertainty due to thermal drifts.

To observe the transverse mode of the OPO (cross-section images shown in Fig. \ref{fig.profile}), we used a CCD camera. The beam was first reflected off a mirror, which was pierced to allow the lattice beam to go through (Fig. 1), and then collimated and refocussed 50~cm later on the camera. This allowed us to image the whole beam, and also to use several small black masks mounted on thin glass plates to get rid of stray reflections of the lattice beam on the vacuum chamber windows, which focused at intermediate distances. The angle calibration used the probe beam, which could also be imaged on the camera, as a reference. The incident angle of the probe beam was precisely determined using a series of Bragg reflection spectra off the passive lattice. When the spectra were symmetric, the probe angle fulfilled $\cos \theta = \lambda_0/\lambda_\dip$ \cite{Schilke:2011}.

\end{scriptsize}

\

\noindent Received 7 July 2011; accepted 14 November 2011; published online 18 December 2011.



\noindent\textbf{Acknowledgments}

\noindent{\scriptsize We acknowledge support from the Alexander von Humboldt foundation, the Deutsche Forschungsgemeinschaft (DFG) and the Research Executive Agency (program COSCALI, No. PIRSES-GA-2010-268717).}

\

\noindent\textbf{Author contributions}

\noindent{\scriptsize A.S. and W.G. performed the experiment and analyzed the data, W.G. supervised the project and wrote the paper. All authors discussed the results and commented on the manuscript.}

\

\noindent\textbf{Additional information}

\noindent{\scriptsize The authors declare no competing financial interests. Supplementary information
accompanies this paper at www.nature.com/naturephotonics. Reprints and permission
information is available online at http://www.nature.com/reprints.
Correspondence and requests for materials should be addressed to W.G.}

\newpage

\onecolumn

\large

\noindent \textbf{Supplementary information}\\

\normalsize

\noindent \textbf{Beat note experiment}\\

To determine the frequency of the emitted radiation, respectively to the pump frequency, we use our probe beam as a local oscillator and perform a beat note experiment.

Both beams are phase-locked together through an injection locking, which allows the resolution of very narrow spectral features. However, both beams pass through different double-pass acousto-optic modulators (AOMs) to allow us to sweep their frequencies. Standard AOM drivers suffer from thermal drifts and thus do not provide a kHz precision. As a consequence, for this beat-note experiment, we use two signal generators (Rohde \& Schwarz SMR20) to drive the AOMs with a well-defined frequency difference of $f_0/2=150$~kHz (because of the double-pass configuration, the frequency difference between both laser is $f_0=300$~kHz).

\begin{figure}[h]
\centerline{\includegraphics[width=7cm]{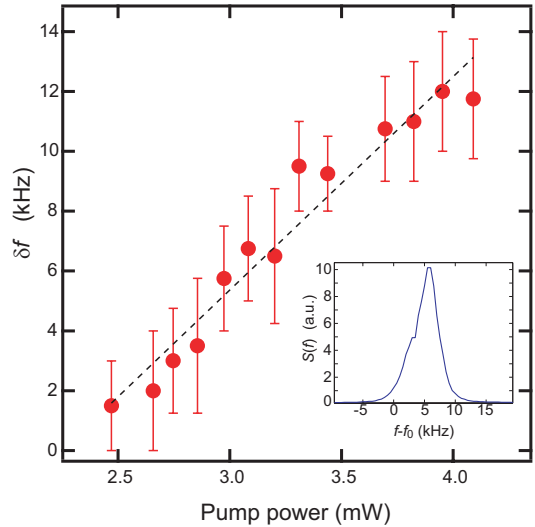}}
\caption{Beat note experiment. Red dots: frequency difference $\delta f$ between the pump and the emitted radiation, as a function of the pump power. We observe a linear frequency drift (dashed line). Inset: example of measured spectrum. The error bars of the main panel correspond to the peak width (FWHM) of the beat-note spectrum.}\label{fig.BeatNote}
\end{figure}

We overlap the probe beam on one of our detection photodiodes and record the temporal trace of the emitted radiation beating with the probe. We generate the Fourier transform of this signal with a digital oscilloscope and we average this power spectrum $S(f)$ over many realizations. We observe a peak around $f_0$ (inset of Fig. S\ref{fig.BeatNote}), whose precise position gives the relative frequency shift $\delta f$ between the pump beam and the emitted radiation. The precision of the measurement is limited by the duration of the OPO.

Even though the most efficient configuration for four-wave mixing is when all waves are degenerate, we observe a non-zero frequency difference. This can be explained by the role of the refractive index, which provides the feedback, and whose value is not necessarily maximum at the pump frequency. The resulting working frequency of the OPO comes thus from a trade-off between the gain efficiency and the feedback efficiency.

We measured the frequency difference $\delta f$ for different pump powers and observe a linear drift (Fig. S\ref{fig.BeatNote}), which is consistent with a broadening of pump-induced spectral structures.
\\

\end{document}